\documentclass[pra,aps,twocolumn,amsmath,amssymb,amsfonts,showpacs]{revtex4}

\usepackage{graphicx}

\newcommand{\be}{\begin{equation}}
\newcommand{\ee}{\end{equation}}
\newcommand{\ba}{\begin{array}}
\newcommand{\ea}{\end{array}}
\newcommand{\bea}{\begin{eqnarray}}
\newcommand{\eea}{\end{eqnarray}}

\newcommand{\bra}[1]{\ensuremath{\langle #1 |}}
\newcommand{\ket}[1]{\ensuremath{| #1 \rangle}}

\newcommand{\ie}{{\it{i.e.~}}}

\begin{document}

\title{Engineering steady states using jump-based feedback for multipartite entanglement generation}

\author{R. N. Stevenson}
\author{J. J. Hope}
\author{A. R. R. Carvalho}
\affiliation{Department of Quantum Sciences, Research School of Physics and Engineering, The Australian National University, ACT 0200, Australia}

\begin{abstract}

We investigate the use of quantum-jump-based feedback to manipulate the stability of multipartite entangled dark states in an open quantum system. Using the model proposed in Phys. Rev. A {\bf 76} 010301(R) (2007) for a pair of atoms, we show a general strategy to produce many-body singlet stationary entangled states for larger number of atoms. In the case of four qubits, we propose a simple local feedback control that, although not optimal, is realistic and stabilises a highly entangled state. We discuss the limitations and analyse alternative strategies within the framework of direct jump feedback schemes.

\end{abstract}
\maketitle

\section{Introduction}

The past two decades have produced impressive progress in the reliable preparation and manipulation of single quantum systems. To achieve this, one of the major challenges has always been the isolation of individual systems and the suppression of decoherence. Over the years, the efforts in this direction moved from simply perfecting technical aspects of experiments that minimise environmental effects, to more elaborated strategies to preserve quantum coherence. Several tactics have been considered in this context, including reservoir engineering~\cite{Poyatos:1996,Carvalho:2001}, dynamical decoupling~\cite{Viola:1999,Viola:2004}, decoherence-free subspaces (DFS)~\cite{Zanardi:1997,Lidar:1998}, open loop control~\cite{Warren:1993} and feedback methods~\cite{Wiseman:1994,Vitali:1998,Doherty:1999}.

Many strategies to control decoherence have been developed recently in the context of quantum information, with a focus on the preparation and preservation of entangled states. Schemes to protect entanglement using reservoir engineering~\cite{Verstraete:2009}, quantum feedback~\cite{Mancini:2005,Mancini:2007,Stockton:2004,Wang:2005,Carvalho:2007,Carvalho:2008,Stevenson:2011}, and reservoir monitoring~\cite{Carvalho:2011} have been proposed, while long-lived entangled states have been produced in the lab using DFS ideas~\cite{Roos:2004,Haffner:2005a}. A remaining challenge in this field is to devise methods to control entanglement in large quantum systems. As the number of components grows, in general the fragility of entanglement against decoherence increases~\cite{Carvalho:2004a,Simon:2002,Dur:2004} and the design of control methods becomes more demanding.

In this paper we address the problem of stabilising multipartite entanglement using direct, or Markovian, feedback based on continuous monitoring of quantum jumps. In this case, feedback is directly proportional to the measurement signal and a master equation representing the unconditioned dynamics of the system can be obtained~\cite{Wiseman:1994}. We show, using an extension of the model proposed in~\cite{Carvalho:2007,Carvalho:2008} for multiple qubits, how the introduction of feedback can change the stability of the steady states of the uncontrolled master equation and be used to protect highly entangled states. In this case, feedback is used to engineer the reservoir by modifying the decoherence operator, which selects preferred steady states of the system. In particular, we show a general strategy to generate and stabilise specific entangled steady states and describe a simple and realistic feedback control for the case of four qubits.

The paper is organised as follows. In Section~\ref{sec:genstruc} we briefly discuss the general structure of the master equation describing a system under jump-based feedback and its consequences to the properties of steady states. In Section~\ref{sec:model} we describe the model where entanglement generation will be investigated. In particular we analyse how the steady state entanglement in the system scales with the number of particles. The effects of feedback in the structure and dynamics of subspaces is presented in Section~\ref{sec:dynamics}. In Section~\ref{Sec:GenStrat} we describe a general strategy to manipulate the steady states of the system to protect specific target states. In Section~\ref{sec:4part} we apply this strategy in the case of four particles and present an alternative simple local feedback to produce stationary entanglement. Section~\ref{sec:conc} finishes with a discussion about the results and future perspectives.

\section{Steady-steate properties of quantum-jump-based feedback}
\label{sec:genstruc}

The evolution of an open quantum system interacting with a single-channel environment can be described, under the assumption of complete positivity and Markovian dynamics, by a master equation in the Lindblad form~\cite{Lindblad:1976,Gorini:1976}:
\begin{equation}
\label{eqn:lindblad}
	\dot\rho = \gamma \mathcal D [\hat c] \hat \rho 
	= \gamma \left[\hat c \hat \rho \hat c^\dagger - \frac{1}{2}\left(\hat c^\dagger \hat c \hat \rho + \hat \rho \hat c^\dagger \hat c\right)\right],
\end{equation}
where $\gamma$ measures the strength of the system-environment coupling and $\hat c$ is the operator that describes the effect of the bath on the system. 

To discuss feedback dynamics, we first need to consider how the system is being measured. Eq.~(\ref{eqn:lindblad}) corresponds to the dynamics of an ensemble of quantum systems described by the density matrix $\rho$ and does not convey any information on how a single quantum system would evolve under measurement. Consider instead the situation where information about the system is extracted by continuously monitoring the environment, more specifically in terms of quantum jumps. The time evolution is then conditioned on the measurement outcomes, and can be described in terms of stochastic quantum trajectories~\cite{Carmichael:1993,Molmer:1993}. The dynamics will intersperse random quantum jumps, defined by the action of the operators $\hat c$, with periods of continuous evolution, given by the non-hermitian effective Hamiltonian $H_{\rm eff}=-i \hbar \gamma \hat c^{\dag} \hat c/2$, corresponding to periods with no detection. 

Now consider that the detection events trigger feedback pulses, which are assumed to be short compared to the other timescales in the system. In this case, the action of the jump operator $\hat c$ is followed by the effect of feedback and the state transforms according to $\ket{\psi} \rightarrow \hat U \hat c \ket{\psi}$, where $\hat U$ is a unitary operator representing the feedback~\cite{Wiseman:1994}. In the absence of detection, no feedback is applied and the system evolves with the previous effective Hamiltonian for the no-jump evolution. 

Because the feedback is assumed to be instantaneous, one can also obtain a Markovian master equation for the jump feedback~\cite{Wiseman:1994}. Following the reasoning above for a single trajectory, one can see that the addition of feedback corresponds to simply replacing the jump $\hat c$ by the combined action $\hat U \hat c$ in the dynamics. The master equation then becomes~\cite{Wiseman:1994}
\begin{equation}
\label{eqn:feedback}
	\dot\rho = \gamma \mathcal D [\hat U\hat c] \hat \rho. 
\end{equation}
Feedback can then be viewed as an instance of reservoir engineering as it allows us to modify the decoherence operator in the master equation through the unitary $\hat U$.  This modified irreversible evolution can then be used to dynamically protect quantum states. The idea is to engineer the steady states of Eq.~(\ref{eqn:feedback}) such that the system naturally evolves towards the desired target state. We therefore need to analyse the stationary states of Eq.~(\ref{eqn:feedback}) and their stability conditions. The mathematical aspects of the stability of steady states of Lindblad master equations have been addressed in the past~\cite{Spohn:1976,Spohn:1977,Frigerio:1978}, including the application in Markovian feedback using homodyne detection~\cite{Schirmer:2010}. Here we will only briefly discuss some of the basic properties that will allow us to construct feedback strategies based on quantum-jump monitoring.  

Note first that the pure steady states, $\ket{\psi_{ss}}$, of Eq.~(\ref{eqn:feedback}) are the eigenstates of $\hat c$ with eigenvalue 0~\footnote{If $\hat c$ is Hermitean then any eigenstate (not only those with eigenvalue 0) would be a steady state of the master equation without feedback. However, any non-trivial control $\hat U$ would break this condition.}. Note also that the introduction of feedback does not affect these states since we also have $\hat U\hat c \ket{\psi_{ss}} = 0$. The dark states are therefore uniquely determined by the structure of the jump operator $\hat c$ and not by the feedback. Although this limits our ability to engineer the dynamics using quantum-jump control, feedback can still be extremely useful to modify the stability properties of the steady states. Indeed, even though the combined jump $\hat U \hat c$ can't move the system once it is in a steady state, it can induce transitions from outside into the dark subspace and be used to select a single dark state to the exclusion of others.

In the remainder of this paper we will analyse this general jump-feedback structure in the context of the model proposed in~\cite{Carvalho:2007} with the goal of generating multipartite entangled steady states.

\section{Model}
\label{sec:model}
 
\begin{figure}
\resizebox{0.9\columnwidth}{!}{
\includegraphics{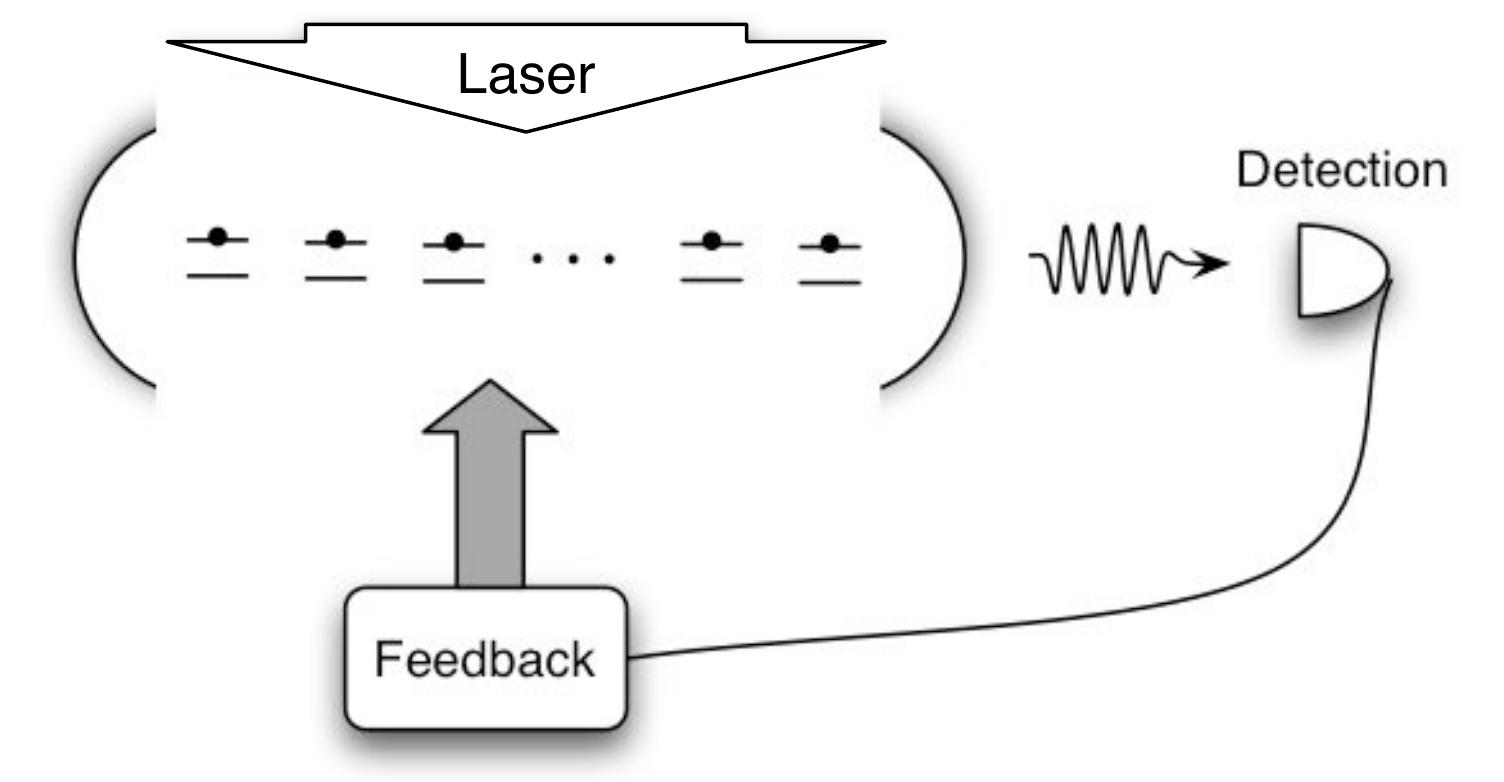} 
}
\caption{Diagram of the model. A series of $N$ two-level atoms are coupled resonanltly to a cavity and equally driven by an external laser field. Light that leaks from the cavity is detected in a photodetector, triggering a feedback pulse.}
\label{Fig:ModelPicture}       
\end{figure}

The system consists of $N$ two-level atoms resonantly driven by a laser with Rabi frequency $\Omega$, and coupled, with strength $g$, to a single cavity mode as depicted in Fig.~\ref{Fig:ModelPicture}. We assume that the light field leaks through one of the cavity mirrors at a rate $\kappa$ and is detected by a photodetector. When the photodetector registers a photon, a short feedback pulse is applied to the atoms. In the limit where the cavity decay rate is large compared to the other rates of the system, we can adiabatically eliminate the cavity mode~\cite{Wang:2005,Carvalho:2008} and obtain the master equation:
\begin{equation}
\label{eqn:model}
\dot \rho=-i\Omega\left[(\hat{J}_+ + \hat{J}_-),\rho \right] + \Gamma {\cal D}[\hat{U} \hat{J}_-]\rho + \sum_j \gamma_j {\cal D}[\sigma_{j,-}]\rho,
\end{equation}
where $\Gamma = g^2/\kappa$ is the effective collective decay rate. The angular momentum operators are defined as $\hat{J}_\pm = \sum_j \sigma_{j,\pm}$, where $\sigma_{j,-}=\ket{g_j}\bra{e_j}$ are the lowering operators for the atomic electronic levels. We also consider the possibility of atomic spontaneous emission with rates $\gamma_j$, as described by the third term in the master equation. At this stage we will not assume any specific form for the feedback Hamiltonian $H_{\rm fb}$, only that its net effect is the action of the unitary evolution 
\begin{equation}
\hat{U} = \mathrm{exp}\left[i \hat H_{\rm fb} \delta t/\hbar\right] \equiv e^{i \hat F}.
\end{equation}

The second term in Eq.(\ref{eqn:model}) is of the form of the feedback master equation described in the previous section. As discussed before, the steady states from this part of the dynamics will be given by the eigenstates of the operator $\hat{J}_-$ with eigenvalue zero. However, these states won't be steady states of the full equation since we have extra Hamiltonian and decay terms. Ignoring the spontaneous emission effects for the moment, we can see that simultaneous eigenstates of $\hat{J}_-$ and $\hat{J}_+$ with eigenvalue zero will not evolve under the Hamiltonian nor under the feedback terms, and will determine the subspace of steady states of the system. In what follows we will describe this subspace structure and the entanglement properties of steady states.

\subsection{Subspace structure}
\label{Sec:SubStruc}

The Hilbert space of a two-level atom is equivalent to that of a spin-$1/2$ system. Our system then consists of $N$ of these pseudo-spins and it is convenient to use an angular momentum basis to describe the problem. The eigenstates of this multipartite system can be found by the usual rules of addition of angular momentum and the relevant quantum numbers are the total spin $J$ and the total spin projection $J_z$. The dynamics in our system is dictated by the operators $\hat J_+$ and $\hat J_-$, whose action on the eigenstates $|J, J_z\rangle$ is given by
\begin{equation}
	\hat{J}_\pm |J, J_z, \lambda_J^k \rangle =  \sqrt{(J \mp J_z)(J \pm J_z+1)} |J,J_z \pm 1 ,\lambda_J^k \rangle.
\end{equation}
Within each total spin subspace (excepting where $J = N/2$) there exist smaller subspaces that are closed under $\hat J_{\pm}$, and these are distinguished by the extra index $\lambda_J^k$. The index $J_z$ determines different states within these subspaces that can be connected through the $\hat J_{\pm}$ operators. The division of the full $J$ subspaces into these smaller subspaces is not unique, but our analysis is independent of the choice of basis that defines this division. Note also that the subspace of pure steady states discussed before corresponds to $J = 0$, a situation only achievable for an even number of particles where $J$ can assume the values $0, 1, ..., \frac{N}{2}$. These states, known as many-body singlets, have interesting entanglement properties~\cite{Toth:2007,Livine:2005,Yao:2011,Toth:2010} with applications to quantum communication~\cite{Bartlett:2003} and computation~\cite{Cabello:2003}.

\subsection{Multipartite entangled steady states}

Having described the subspace of pure steady states, we are now in position to investigate the entanglement properties of all dark states within this subspace. To do this we will use the $C_N$ concurrence~\cite{Carvalho:2004a,Mintert:2005} as a measure of entanglement in the system. Although the correlations in a $N$-partite system cannot be completely characterised by a single scalar quantity, the $C_N$ concurrence is useful to analyse the scaling properties of entanglement~\cite{Carvalho:2004a}. 

\begin{figure}
\resizebox{0.9\columnwidth}{!}{
  \includegraphics{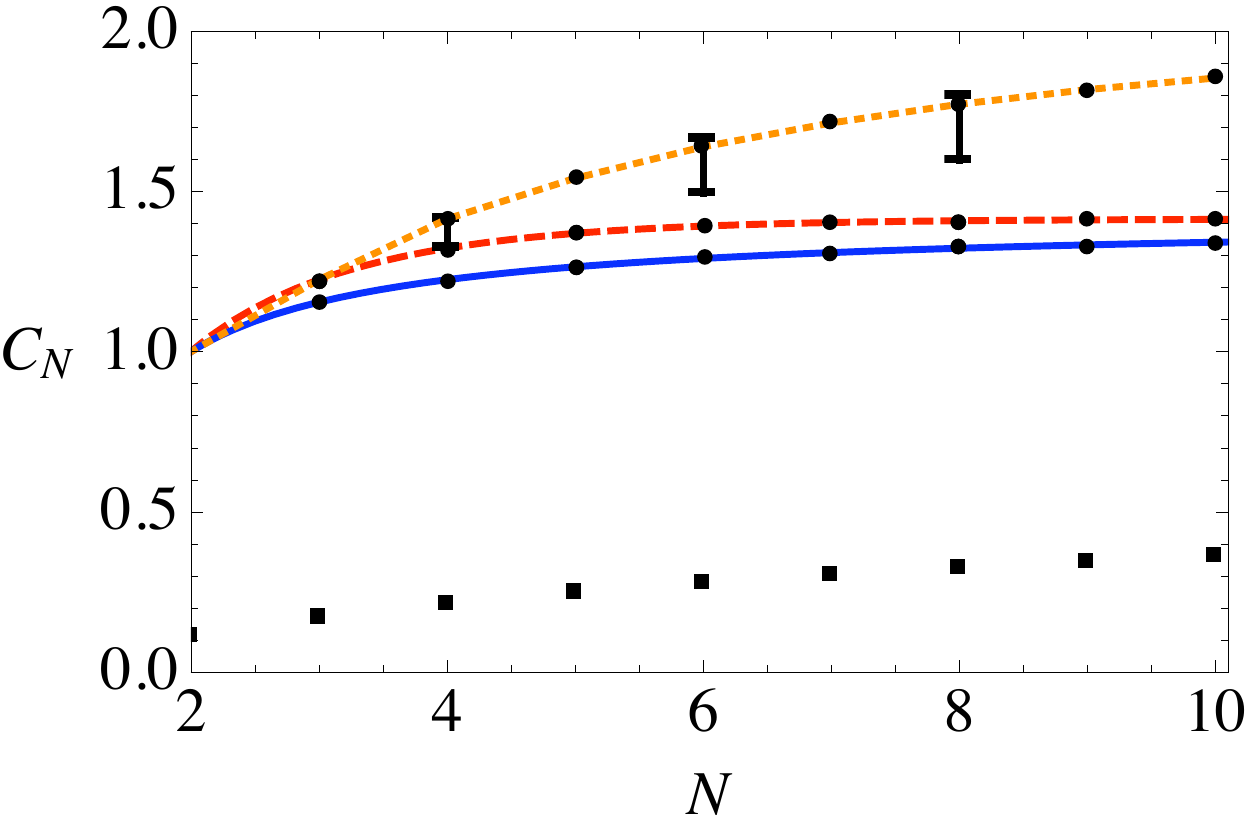} 
}
\caption{$N$-concurrence for various kinds of states as a function of the number of atoms. Vertical bars indicate the range of the concurrence for the many-body singlets forming the subspace of pure steady states of Eq.~(\ref{eqn:model}) with $\gamma_i=0$. Concurrence for GHZ, W and linear cluster states are shown by dashed, solid, and dotted lines, respectively. Solid squares represent the stationary states of Eq.~(\ref{eqn:model}) in the absence of feedback, when all atoms start in the ground state.}
\label{Fig:ConcurrencePicture}       
\end{figure}

Let us start with the example of a 4-partite system. In this case the subspace of dark states can be parametrised as
\begin{eqnarray}
 |\psi_{\rm ss}\rangle&=&\frac{1}{\sqrt{\cal N}}\left[\alpha  e^{i \theta } |ggee\rangle +\beta  e^{i \phi } |gege\rangle \right.\nonumber \\
 &-&\left(\alpha e^{i \theta }+\beta  e^{i \phi }\right)|geeg\rangle  -\left(\alpha e^{i \theta }+\beta  e^{i \phi
   }\right)|egge\rangle \nonumber \\ &+&\beta  e^{i \phi } |egeg\rangle+\alpha  e^{i \theta } |eegg\rangle \left. \right],
\label{eqn:states}
\end{eqnarray} 
where $\cal N$ is a normalisation factor. The $N$-concurrence of elements of this set varies from $\sqrt{7}/2$ to $\sqrt{2}$, depending on the values of $\alpha$, $\beta$, $\theta$ and $\phi$. This range is plotted in Fig.~\ref{Fig:ConcurrencePicture} together with the equivalent range for $N=6$ and $N=8$ atoms. Curves for representative states like GHZ, W, and linear-cluster states~\cite{Briegel:2001} are also displayed for comparison. These results show that, in the absence of spontaneous emission, the model described by Eq.~(\ref{eqn:model}) has highly entangled steady states
, but it does not automatically follow that any initial state will lead to a highly entangled steady state.  In the absence of feedback, each component of the initial state with a well-defined angular momentum stays within that subspace.  If the system starts in a dark state then it will remain in that state indefinitely. However, if the initial state has a little overlap with the pure dark states, then the system will likely evolve towards a mixed steady state with low entanglement. This situation is represented by the filled squares in Fig.~\ref{Fig:ConcurrencePicture}, which correspond to the steady state concurrence achieved after evolving the system from all atoms starting in the ground state. This shows that despite the existence of pure steady states, these are not attractive points of the dynamics and therefore the system does not evolve naturally towards them. Our goal here is to design a feedback that could change this scenario by making the system dynamically evolve to a target highly entangled state.

\section{Dynamics between subspaces in terms of quantum trajectories}
\label{sec:dynamics}

The feedback design relies on understanding the detailed dynamics of the system and its consequences to the structure and stability of steady states. A natural way to approach this problem is to use the quantum trajectory method described briefly in Section~\ref{sec:genstruc}. The method not only provides a faithful picture of what a single run of an experiment would look like, but also provides a sensible splitting of the dynamics in terms of jumps, when control acts, and a continuous evolution independent of the feedback.

\subsection{The action of jumps}

When a detection event occurs, the state of the system changes with the application of the combined jump operator $\hat U \hat J_-$:
\begin{equation}
\ket{\psi^{\prime}}=\frac{\hat U \hat J_- \ket{\psi}}{\sqrt{\langle \hat J_+ \hat J_- \rangle}}.
\end{equation} 
The $\hat J_-$ operator does not mix subspaces with different $J$'s but rather induces transitions towards the states with minimum spin projection ($J_z=-J$) within each subspace. However, the subsequent application of the feedback unitary can cause transitions between different subspaces and be used to help move the system to a target state. This can happen, for example, if the intermediate state $\ket{\psi_i}$, which results from the action of $\hat J_-$ on the initial state $\ket{\psi_0}$, is connected to the targeted steady state $\ket{\psi_{\rm t}}$ via the feedback operator $\hat U$, \ie,  
\begin{equation}
|\psi_{\rm t}\rangle=\hat U |\psi_i\rangle=\hat U \hat J_- |\psi_0\rangle, 
\end{equation}
where normalisation factors have been omitted for simplicity. Once the system is fully in the steady state then no more jumps will happen since $\hat U \hat J_- |\psi_{\rm t}\rangle = 0$. 

Note, however, that the jump dynamics is not enough to account for a full transfer of population to one of the pure steady states of the system. First, because states with $J_z = -J$ are also dark to the action of $\hat U \hat J_-$ and any initial population in those states cannot be directly connected only through jump dynamics to the target states.
Second, because if the system is not entirely in the dark subspace, then a jump event will have exactly the opposite of the desired effect. Consider, for example, the system initially in a mixed state $\hat \rho= \vert\alpha\vert^2 \ket{\phi}\bra{\phi} + \vert\beta\vert^2 \ket{\psi_{\rm t}}\bra{\psi_{\rm t}}$, where $\ket{\phi}$ is not in the subspace of steady states. Then, the action of the operator $\hat U \hat J_-$ will lead to the final pure state $\ket{\psi^{\prime}}= \hat U \hat J_- \ket{\phi}$ and, unless $\ket{\phi}$ is one of the states connected to the target state via the jump operation, then the system will be completely removed from the dark subspace. This process, however, has a probability to occur that goes to zero as $\beta\rightarrow 1$.

\subsection{Interplay between jump and no-jump evolutions}

The dynamics between detection events is given by 
\begin{align}
    \frac{d}{dt}|\psi \rangle &= -\frac{i}{\hbar} (\hat H + \hat H_\text{eff}) |\psi\rangle \\
                        &=- i \Omega (\hat J_+ + \hat J_-) |\psi\rangle - \Gamma \hat J_+ \hat J_- |\psi\rangle,
\end{align}
where the first term represents the driving Hamiltonian of Eq.~(\ref{eqn:model}), and the second term is the non-Hermitian effective evolution described in Section~\ref{sec:genstruc}. The solution at any given time can be found by integrating this equation and then renormalising the state. This last step, necessary due to the norm decay induced by the effective Hamiltonian, is crucial to understand the transfer of population in the absence of detections.  

The norm of different components of the state will decay at different rates, and renormalisation will transfer population from states that decay quickly to states that decay slowly. Physically, this can be understood from the dynamics of the system conditioned on the measurement results. As we continuously monitor the system and no jump events are observed, we acquire information that it is more probable that the system is in a state where jumps are less likely to occur. To make this statement more quantitative, let us look at the action of the effective Hamiltonian on our basis states: 
\begin{align}
\hat J_+ \hat J_- |J, J_z, \lambda_J^k\rangle = (J + J_z)(J - J_z + 1)|J, J_z, \lambda_J^k\rangle
\end{align}
The norm decay will be proportional to the factor $(J + J_z)(J - J_z + 1)$, which is zero for states with $J_z = -J$ and is at least equal to $2 J$ for all other eigenstates. This means that the renormalisation step will gradually transfer the population not only to the dark subspace with $J=0$, but to all the states with the lowest spin projection ($J_z = -J$).

The remaining component of the no-jump dynamics is the driving Hamiltonian. This is the term that ensures that only the population in the steady state subspace will increase. Indeed, it affects all states with $J \not = 0$, transferring population from the $J_z = -J \not = 0$ into states with norm decay, while leaving the subspace of steady states unaffected. Therefore, the net effect of the dynamics between detection events is to increase the population in the dark subspace.

However, this dynamics is not sufficient to protect a particular highly entangled steady state of the system. As described in Section~\ref{Sec:SubStruc}, the dark subspace of the dynamics has multiple linearly independent states. These states evolve exactly in the same way under the no-jump evolution, which will not be able to select one of these states over the others, as it can for states with $J\ne 0$. The fact that the system can be in different dark states in the quantum trajectory picture means that in the full density matrix description the state is in a mixed state, which has lower entanglement than the pure dark states. To ensure that only one target state in the subspace of steady states is accessed, we need to make sure that the system can't reach the other orthogonal states in this subspace. 

\section{General feedback strategy}
\label{Sec:GenStrat}

A general strategy for achieving a pure steady state can now be formulated. We need to find a feedback unitary $\hat U$ that achieves the two tasks described above: i) eliminating the unwanted states in the dark subspace, and ii) providing transitions between states so that there is a path to the desired state. This will guarantee that the system will dynamically evolve towards a single steady state from any non-steady initial state. 

To design a feedback fulfilling these requirements let us first chose a target state from the set of steady states $|J = 0, J_z = 0, \lambda_0^k\rangle$. We will label this state by $\lambda_0^{\rm t}$. All the states orthogonal to the target state in this subspace will then form the set of unwanted dark states $\ket{\psi_{\rm u}}\equiv |J = 0, J_z = 0, \lambda_0^k \ne \lambda_0^{\rm t}\rangle$. These unwanted states can only be accessed from other subspaces via the combined jump operator $\hat U \hat J_-$. A way to avoid these transitions is to have the feedback unitary $\hat U$ connecting to $\ket{\psi_{\rm u}}$ from a state $\ket{\psi_i}$ that cannot be reached through $\hat J_-$, \ie,
\begin{equation}
\ket{\psi} \overset{\hat J_-}{\nrightarrow} \ket{\psi_i} \overset{\hat U}{\rightarrow} \ket{\psi_{\rm u}}, 
\end{equation} 
for any state $\ket{\psi}$. The most obvious states for this are the unwanted dark states themselves so that $\hat U |\psi_{\rm u}\rangle = |\psi_{\rm u}\rangle$, although any state with $J_z = +J$ would do (as there are no states for which $\hat J_- |\psi\rangle$ has $J_z = +J$). As a result, all other matrix elements in $\hat U$ connecting to the unwanted states will be 0: 
\begin{align}
    \langle \psi_{\rm u}|\hat U |J_i, J_z \ne +J_i, \lambda_{J_i}^{k}\rangle = 0
\end{align}
for all $i$.

\begin{figure}
\resizebox{0.8\columnwidth}{!}{
  \includegraphics{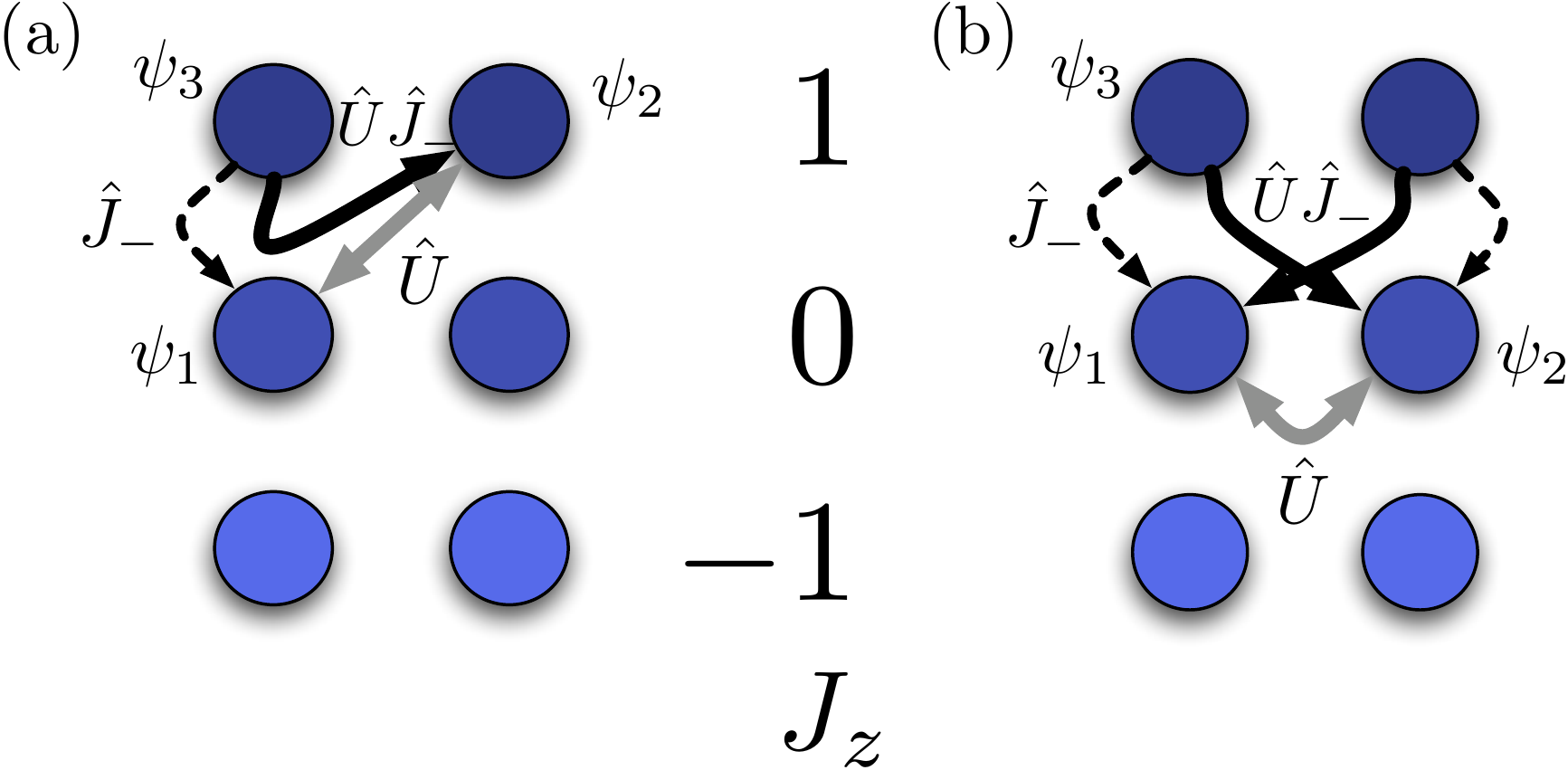}
}
\caption{Transitions between subspaces (columns) induced by the feedback operation. The grey doublepointed arrow shows the interaction via the unitary $\hat U$, the dotted arrow shows the effect of the $\hat J_-$ operator, while the solid black arrow shows the net effect of the combined $\hat U \hat J_-$ operation. Diagrams show transformations that results in jumps that can go from the left to the right subspace (a) (one-way transitions), or in both ways (b) (two-way transitions).}
\label{Fig:connections}       
\end{figure}

The second task is accomplished by engineering a $\hat U$ that mixes all the subspaces with $J \not= 0$. This unitary is such that the jump operator $\hat U \hat J_-$ induces connections between the different $J$ subspaces and provides to each one of them a path to the target state. Figure~\ref{Fig:connections} shows different schemes to establish these connections. Suppose that a state $|\psi_1\rangle$ in one space and $|\psi_2\rangle$ in another are such that $\langle \psi_1 |\hat U |\psi_2 \rangle \ne 0$ (double-arrowed connection in Fig.~\ref{Fig:connections}). If $|\psi_1 \rangle$ has $J_z \not = +J$ then there exists a state $|\psi_3\rangle$ in the same subspace such that $\hat J_- |\psi_3\rangle = |\psi_1\rangle$ (dashed arrow), so that $\hat U \hat J_- |\psi_3\rangle = |\psi_2\rangle$ (solid arrow). Under the action of jumps, the system can then transition from the first subspace to the second. If $|\psi_2 \rangle$ also has $J_z \not = +J$, then the transition can happen in both directions (as shown by the solid arrows in Fig.~\ref{Fig:connections}-b), otherwise the transition is unidirectional as depicted in Fig.~\ref{Fig:connections}-a. The path towards the target state can then be built by a chain of similar connections between subspaces. 

The general strategy presented in this section has a lot of freedom, with a variety of possible feedback operators fulfilling the requirements for stabilisation of a single entangled steady state. These choices may differ, however, on many other aspects such as the time it takes to reach the steady state, the robustness against imperfections or natural decoherence processes, and also on how simple it would be to experimentally implement them. We will leave these issues to the next section where we discuss the concrete case of controlling a system with four particles.

\section{Example: four-partite system}
\label{sec:4part}

\begin{figure}
\resizebox{0.8\columnwidth}{!}{
  \includegraphics{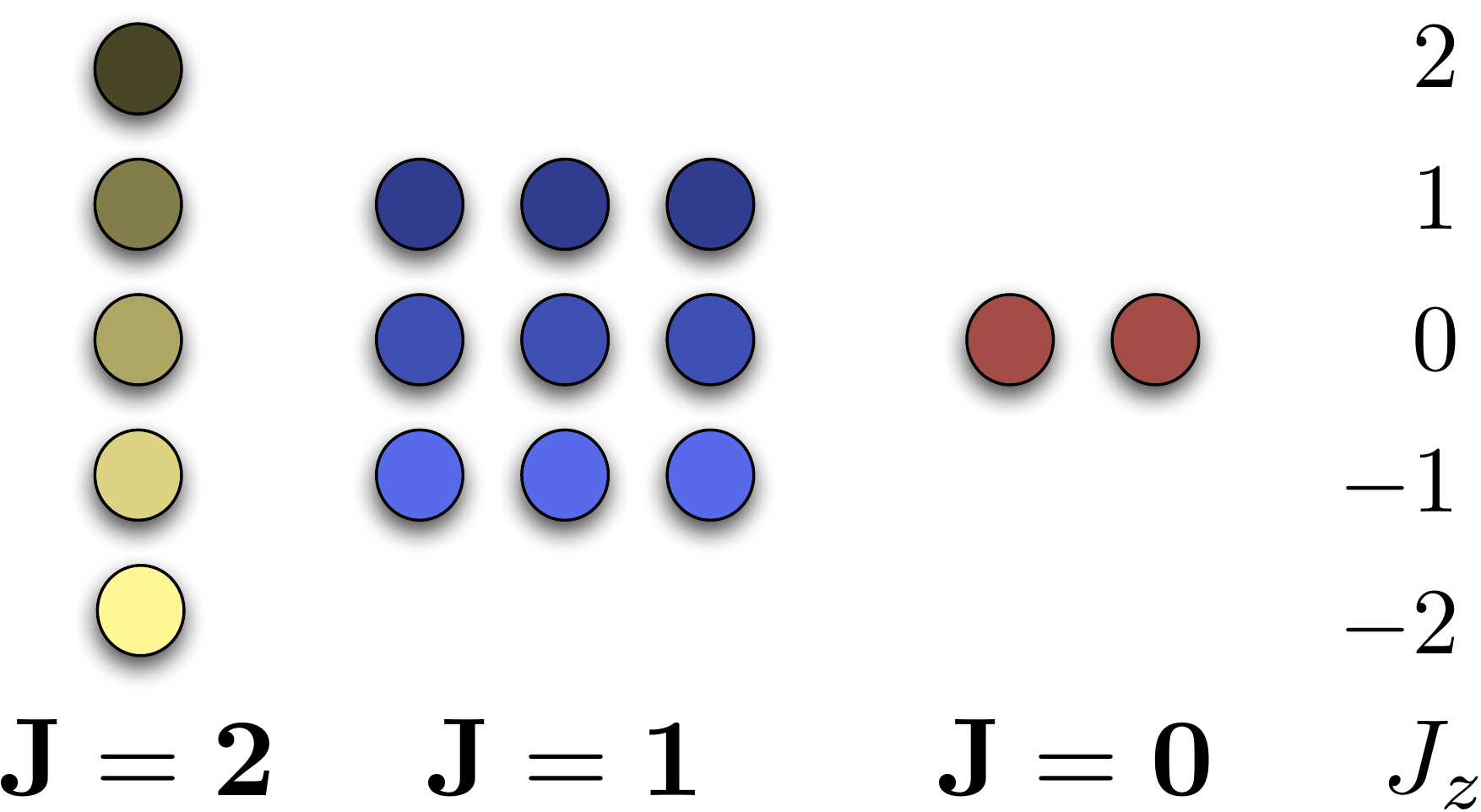}
}
\caption{Diagram of the angular momentum basis states for a four-partite system. States in the same $J$ subspace are grouped together with the vertical ordering following the different values of $J_z$. Different columns within the same $J$ subspace are labelled with different $\lambda_J^k$.}
\label{Fig:StatesPicture}       
\end{figure}

The subspace structure for a 4-partite system is shown in Fig.~\ref{Fig:StatesPicture}. States (represented by circles) within the same $J$ subspace are grouped together, with the vertical position indicating the different $J_z$ projections. Although the configuration in this picture is general, there is some freedom in the basis such that the choice of specific states associated to each circle is arbitrary. If we choose two basis states $\ket{\psi_1}$ and $\ket{\psi_2}$ for the $J = 0$ subspace, for example, then any linear combination of these states will give a different valid choice of basis. 

\begin{figure}
\begin{center}
\resizebox{0.8\columnwidth}{!}{%
  \includegraphics{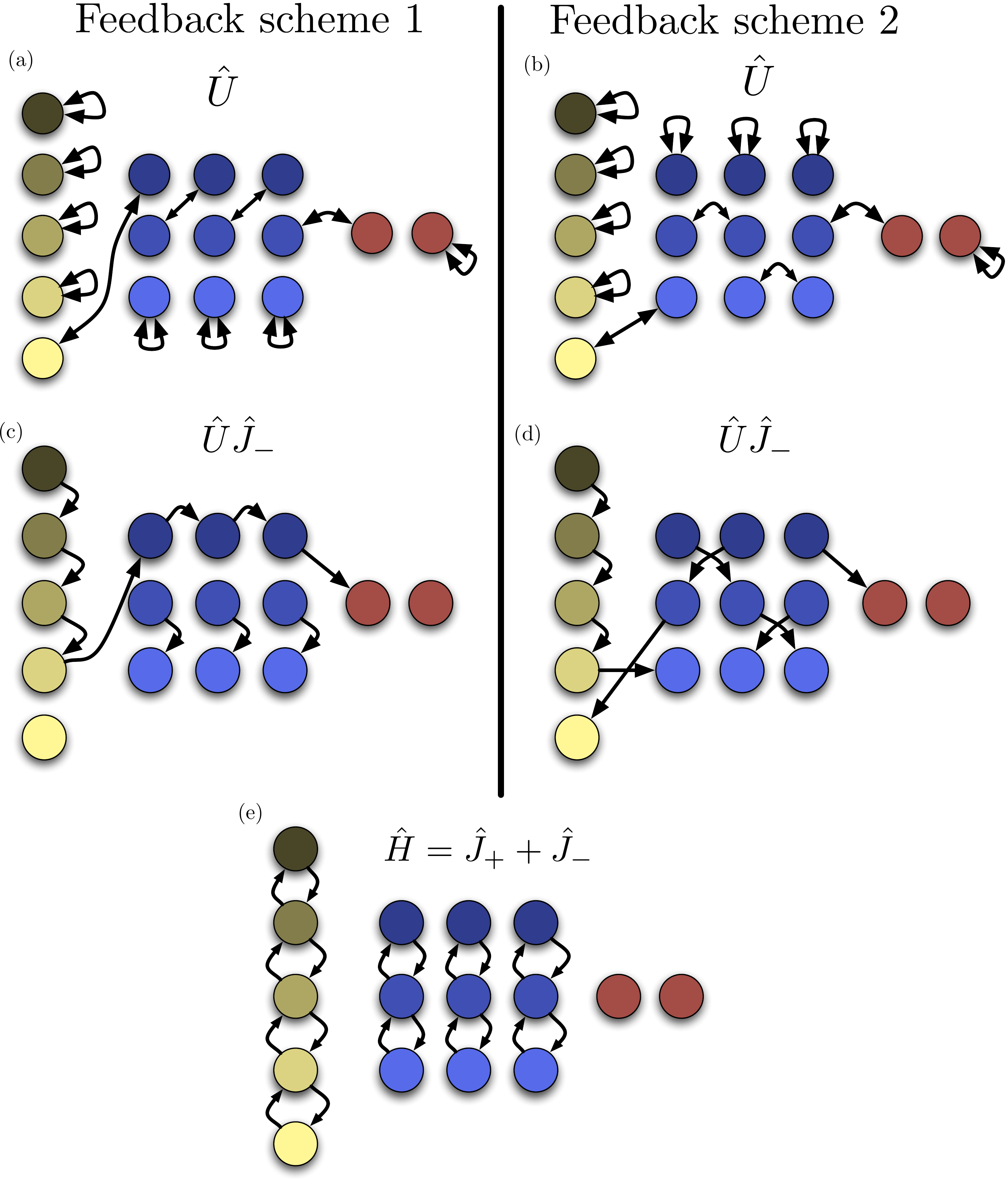}
}
\end{center}
\caption{Schematics of two different feedback schemes.  Diagrams (a) and (b) show the connections between states induced by $\hat U$ while (c) and (d) represent the effect of $\hat U\hat J_-$. The coupling given by the driving Hamiltonian is schematically shown in (e).  One scheme (a and c) has one-way transitions between $J$ subspaces, and the other (b and d) has two-way transitions that still provide a path to the target state. }
\label{Fig:4partU}       
\end{figure}

This schematic representation of the subspaces is very useful to analyse the effect of feedback. Two choices of $\hat U$ that follow the strategy described in Section~\ref{Sec:GenStrat} are shown in Figure~\ref{Fig:4partU} (a and b), along with a representation of $\hat U\hat J_-$ for each (c and d). Connecting arrows represent pairs of states that are coupled through the operators shown in the figure. One feedback uses two-way transitions that don't involve states with $J_z = +J$ (b and d in Fig.~\ref{Fig:4partU}), and the other involves one-way connections in a path from the space with $J = 2$ through to the steady state (a and c in Fig.~\ref{Fig:4partU}). Within each invariant irreducible subspace $\{ J, \lambda_J^k\}$ the different spin projection states are mixed through the Hamiltonian term, and this is shown in Fig.~\ref{Fig:4partU}-e.

In both schemes we can see the requirements described in Section~\ref{Sec:GenStrat} fulfilled: The feedback unitary connecting the unwanted dark state $\psi_{\rm u}$ to itself such that it can't be populated from other states, and a path connecting the different subspaces to the target state $\psi_{\rm t}$. In the direct one-way scheme of Fig.~\ref{Fig:4partU}-c the path towards the target state is obvious and the Hamiltonian interaction depicted in Fig.~\ref{Fig:4partU}-e ensures that the system will not be trapped in any of the $J_z=-J$ states (lowest circles in the $J=2$ and $J=1$ subspaces).

\begin{figure}
\begin{center}
\resizebox{0.8\columnwidth}{!}{%
  \includegraphics{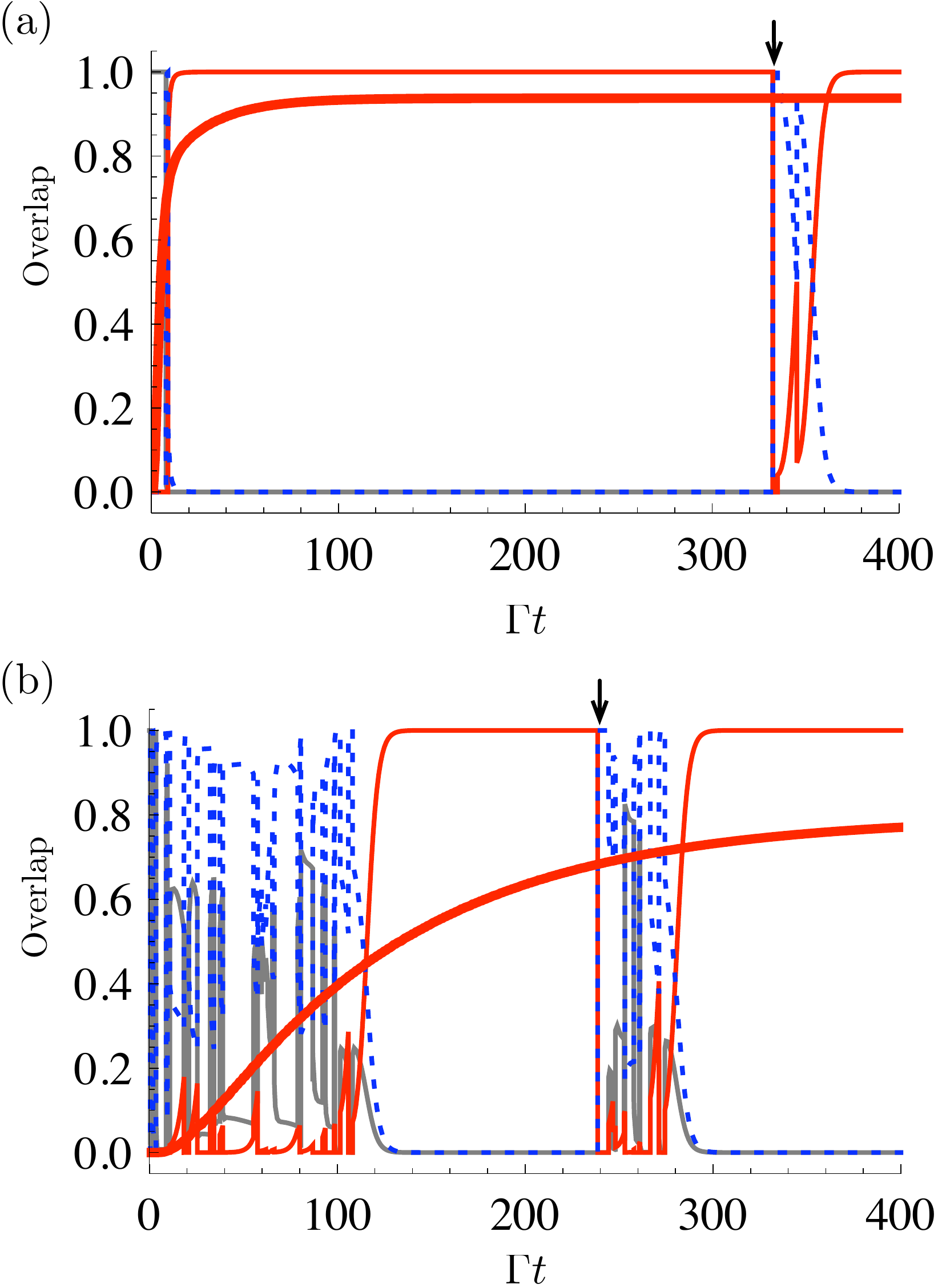} 
}
\end{center}
\caption{Simulations showing the effectiveness of the two feedbacks shown in Figure \ref{Fig:4partU} with a spontaneous emission rate of $10^{-3} \Gamma$. Each curve shows the overlap of the system state with a set of basis states. The thin lines represent the result of a single trajectory simulation and show the overlap with $J = 2$ (solid grey line), $J = 1$ (dotted blue line) and the target steady state (solid red line). The thick red line shows the overlap with the target steady state for a full master equation simulations. The one-way feedback (a) takes fewer jumps and shorter time to reach the steady state than the two-way feedback (b). The dips indicated by the arrows in each of these simulations represent spontaneous emission events.}
\label{Fig:USims}       
\end{figure}

Despite leading to exactly the same final target state, these two choices of feedback are radically different dynamically. Feedback schemes with unidirectional connections are more efficient, in the sense that they bring the system to the target state faster. Indeed, as the system moves from the $J=2$ to the $J=1$ subspace via a jump process, it can't go back. From the $J=1$, a detection event followed by the feedback application can only move the system within the subspace or to the dark ($J=0$) subspace. However, in the case of bidirectional connections, once in the $J=1$ subspace the system can either follow the path towards the $J=0$ or go back to the $J=2$ subspace. This difference in efficiency can be seen in Figure~\ref{Fig:USims}, where the overlap between the state of the system and the target state as a function of time is plotted for a single quantum trajectory. In the unidirectional case (Fig.~\ref{Fig:USims}-a) the system reaches the steady state, on average, ten times faster and with ten times less jumps than the two-way feedback showed in Fig.~\ref{Fig:USims}-b.

The issue of time efficiency turns out to be of crucial importance when a natural decoherence effect, such as spontaneous emission, is taken into account. Spontaneous emission is described by the third term in Eq.~(\ref{eqn:model}) and a single decay event (indicated by arrows in Fig.~\ref{Fig:USims}) will knock the system out of the dark subspace. After the emission, the feedback control dynamics will have to restart the process of pushing the system to the target state. The effectiveness of the control then becomes a matter of competing feedback and spontaneous emission rates. In the simulations shown in Fig.~\ref{Fig:USims}, this means that the unidirectional feedback will be not only faster than the two-way feedback to recover after a spontaneous emission event, but also more effective as the systems will spend more time in the target steady state. The consequence of that for the unconditioned evolution, \ie the solution of the master equation for the density matrix of the system, is that the entanglement of the system with spontaneous emission will be closer to the ideal value for the unidirectional feedback. This behaviour can be seen from the overlap of the unconditioned state $\rho(t)$ with the target state shown by the smooth bold lines in Fig.~\ref{Fig:USims}.

\subsection{Simple implementations of the feedback}
\label{Sec:Local}

In the examples presented above, we discussed how to construct feedback unitaries that satisfy the conditions for state preparation and protection, their efficiency and efficacy  
against spontaneous emission. So far we haven't considered the specific form of these feedback operations in terms of physical processes happening to the atoms. In principle, a generic $\hat U$ obeying our feedback strategy can be quite complicated, even involving multiple qubit interactions. Given that in real experiments one would have limitations to produce a feedback unitary $\hat U$, in this Section we will change the focus and ask a different question: can we find feedback forms that satisfy (maybe only approximately) our general requirements and that are yet simple to implement? We will answer this question by providing an example of a feedback that is based only on single atom operations and that is able to approximately protect one type of entangled state in the dark subspace of the problem. 

The steady state subspace of the four-partite system includes any state that is the tensor product of two anti-symetric Bell states as, for example,
\begin{align}
    |\psi_{BB}\rangle &= \frac{1}{\sqrt{2}} \left(|ge\rangle_{12} - |eg\rangle_{12}\right)\otimes\frac{1}{\sqrt{2}} \left(|ge\rangle_{34} - |eg\rangle_{34}\right) \nonumber \\
    &= \frac{1}{2} \left(|gege\rangle -|geeg\rangle-|egge\rangle+|egeg\rangle\right).
\end{align}
This state corresponds to the choice $\alpha=\theta=\phi=0$ and $\beta=1/2$ in Eq.~(\ref{eqn:states}), and this is the state we can eliminate from the dynamics. As these antisymmetric Bell states are eigenstates of the two-partite total spin with $J_{12} = J_{34}= 0$, they are dark to any local operators that are symmetric over the exchange of the two particles in each pair. Let's consider a feedback Hamiltonian of the form
\begin{equation}
    \hat F = \sum_{i=1}^4 a_i \hat\sigma^x_1 ,
\label{eqn:localfeedback}
\end{equation}
which corresponds to driving each atom with different intensities. If the coefficients $a_i$ are chosen such that $a_1 = a_2$ and $a_3 = a_4$, then $\hat F |\psi_{BB}\rangle = 0$ and, consequently, we have $\hat U |\psi_{BB}\rangle = e^{i \hat F} |\psi_{BB}\rangle=|\psi_{BB}\rangle$, as required for the elimination of unwanted steady states described in Section~\ref{Sec:GenStrat}.

The target state will now be the state with $J = 0$ that is orthogonal to $|\psi_{BB}\rangle$. This state in this case is 
\begin{eqnarray}
\label{eqn:target}
    |\psi_{\rm t}\rangle = \frac{1}{\sqrt{12}} \left( 2 |ggee\rangle + 2 |eegg\rangle - |gege\rangle  \right.\nonumber \\ -|geeg\rangle-|egge\rangle - |egeg\rangle \left. \right),
\end{eqnarray}
which is a highly entangled 4-qubit singlet~\cite{Cabello:2003}.

\begin{figure}
\begin{center}
\resizebox{0.8\columnwidth}{!}{%
  \includegraphics{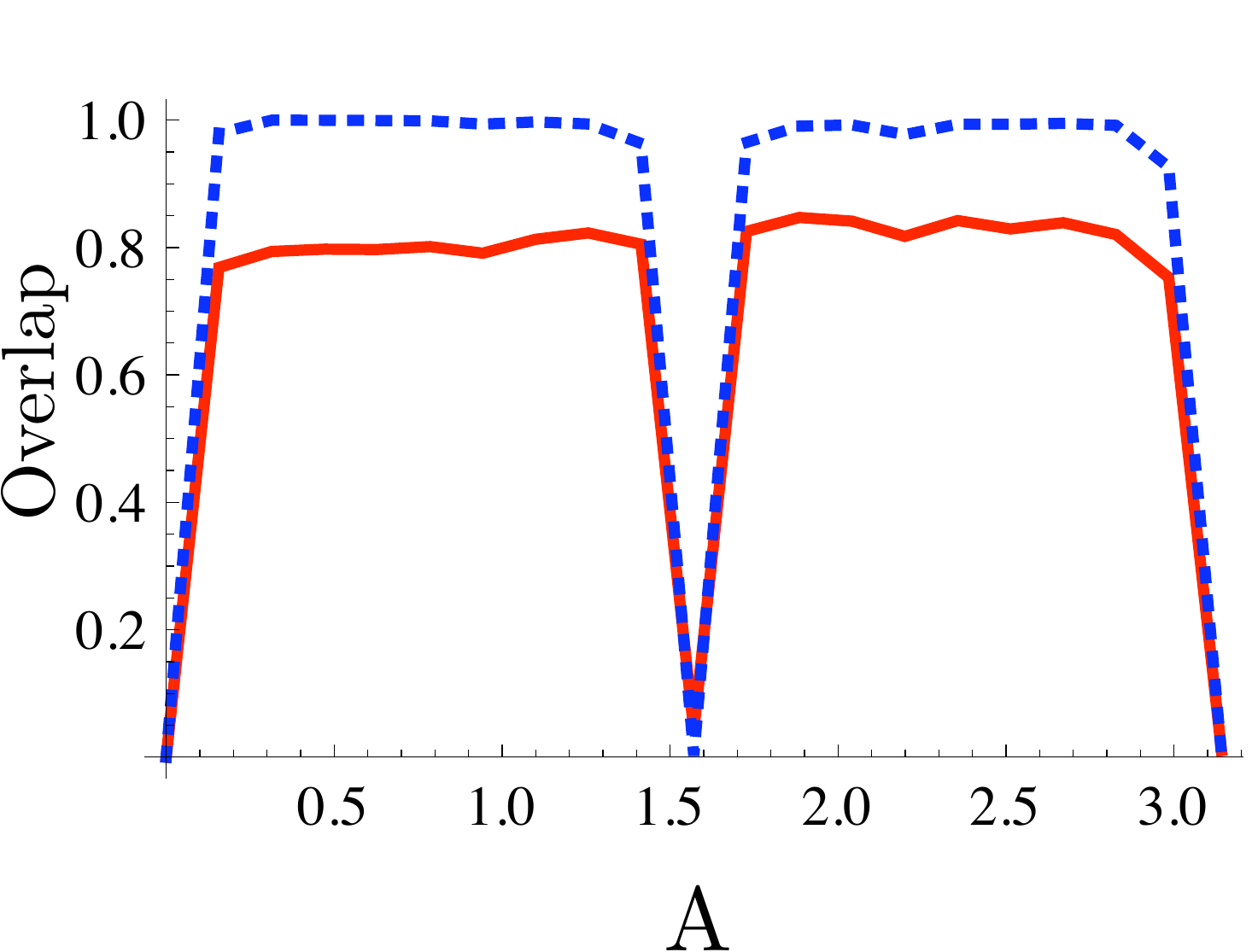} 
}
\end{center}
\caption{Overlap between the final state and the target steady state for the feedback of Eq.~(\ref{eqn:localfeedback}) with $a_1 = a_2 =A$ and $a_3 = a_4 = -A$ as a function of A. The simulations show results without spontaneous emission (dotted blue line) and with $\gamma=\Gamma \times 10^{-3}$ (solid red line). For both of these simulations the system started with all atoms in the ground state. Spontaneous emission induces transitions to trapped states that do not connect to the target state, considerably decreasing the overlap.}
\label{Fig:OverlapSlice}       
\end{figure}

Figure~\ref{Fig:OverlapSlice} shows the overlap of the steady state with the target state $|\psi_{\rm t}\rangle$ for the choice $a_1 = a_2 =A$ and $a_3 = a_4 = -A$, for various values of $A$. Without spontaneous emission (dashed line) the system goes to the steady state (excluding the values $A=k\,\pi$ with $k$ integer). However, when spontaneous emission is taken into account the overlap is less than $80\%$. The reason for that is that our choice of $a_i$'s only achieves the first part of the strategy: eliminating all but one state in the dark subspace. A feedback of this form does not achieve the second goal of mixing the other states, as it leaves some states in the $J = 1$ subspace closed under the dynamics. Without spontaneous emission the mixing of all subspaces is less important since the system will only occupy an isolated subspace if the initial state of the system (in our case $\ket{gggg}$) has a component in this space. Spontaneous emission, however, promotes transitions to states that are not coupled via feedback and the system can be trapped in those states, hence not reaching the final target. Note that with spontaneous emission the initial condition is not important as there can be only one steady state~\cite{Schirmer:2010}.

\begin{figure}
\begin{center}
\resizebox{0.8\columnwidth}{!}{%
  \includegraphics{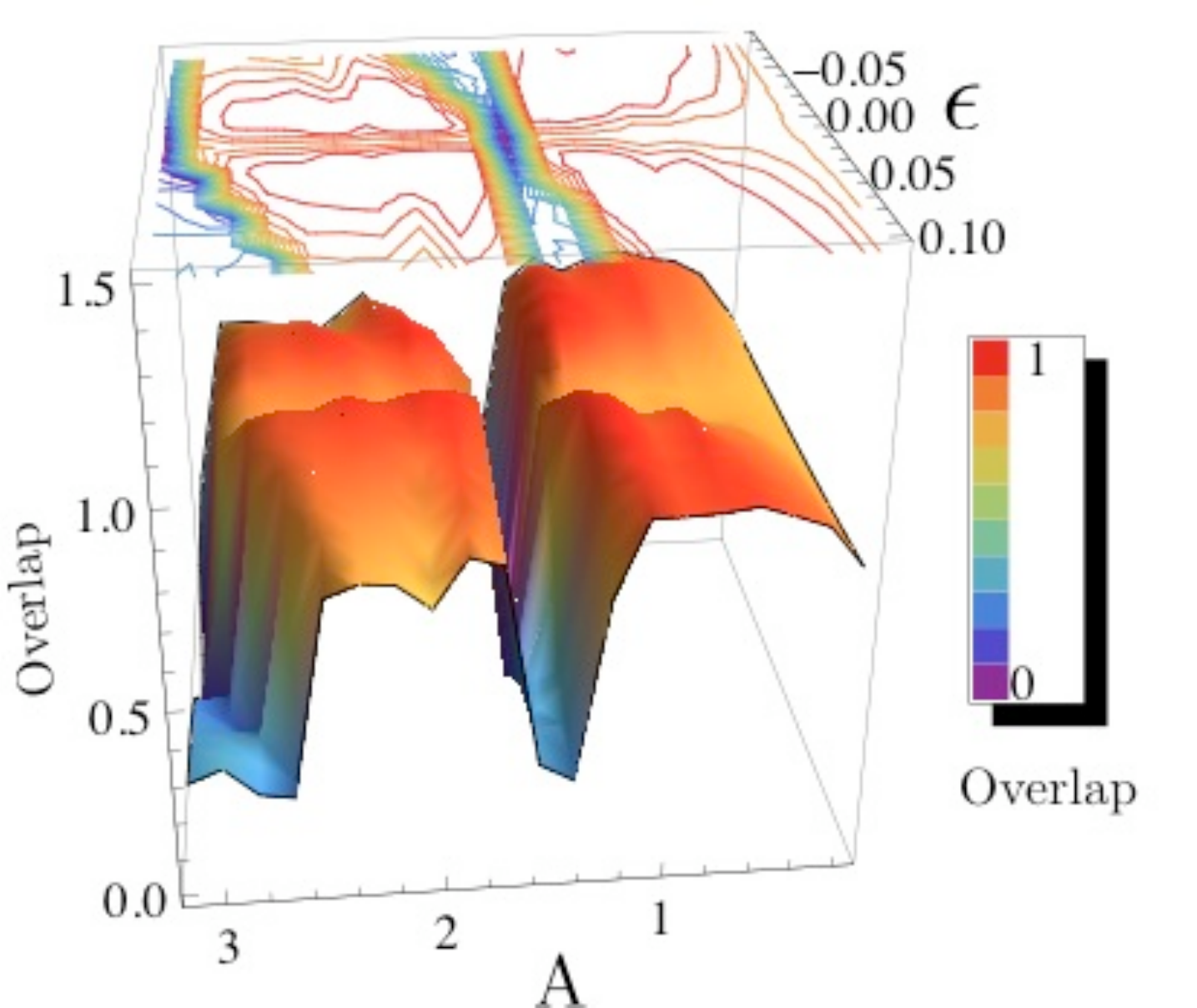} 
}
\end{center}
\caption{Overlap between the final state and the target steady state for the feedback in Eq.~(\ref{Eqn:Local}) as a function of $A$ and $\epsilon$. The spontaneous emission rates for these simulations is $\gamma=\Gamma \times 10^{-3}$. As in Fig.~\ref{Fig:OverlapSlice}, there are dips in the overlap when $A = k \pi$. For $\epsilon \ne 0$ the overlap gets closer to one as compared to the curve in Fig.~\ref{Fig:OverlapSlice}.}
\label{Fig:local}       
\end{figure}

To circumvent this problem and allow a mixing of all the other subspaces, we introduce a small deviation from the conditions $a_1 = a_2$ and $a_3 = a_4$ by considering the following feedback Hamiltonian:
\begin{equation}
    \hat F = A \hat\sigma^x_1 + A (1 - \epsilon) \hat\sigma^x_2 - A \hat\sigma^x_3  - A (1 - \epsilon) \hat\sigma^x_4.
    \label{Eqn:Local}
\end{equation}
Figure~\ref{Fig:local} shows the overlap between the final steady state and the target state as a function of $A$ and $\epsilon$ with the spontaneous emission rates $\gamma_i=\gamma= 10^{-3} \Gamma$. We see that the introduction of the parameter $\epsilon$ is enough to connect the subspaces that were previously isolated mitigating, in this way, the effects of spontaneous emission. For $\epsilon \ne 0$ and $A \ne k \pi$ the overlap is close to unity, demonstrating the efficiency of this particular local feedback to generate the many-body singlet of Eq.~(\ref{eqn:target}).

\section{Conclusions}
\label{sec:conc}

In this paper we have shown how pure entangled states of a system of multiple atoms in a cavity can be prepared and protected using quantum-jump-based feedback. We have formulated a general strategy that explores the steady state properties of the master equation and how they can be modified via feedback control. This engineered evolution dynamically steers the system towards a unique target steady state, inducing, in this way, a robust generation of multipartite entangled states.

The general requirements for an effective feedback presented here are independent of the number of atoms in the system. For this reason, the feedback operators satisfying these conditions can, in principle, be constructed for any number of qubits. However, practical considerations would impose extra constraints on the form of the unitary $\hat U$ that could be reasonably implemented in real experiments. Here we partially addressed this issue by showing that a simple feedback based on single-atom operations can be used to stabilise a given class of multipartite entangled states in the case of four qubits. This example, together with  the freedom in the choice of the feedback $\hat U$, indicate that it may be possible to find equally simple feedback Hamiltonians for systems with arbitrary number of constituents. At the moment, a general approach to this problem remains an open question. 

Finally, we should mention that the dark subspace cannot be changed by the feedback as it is solely determined by the measurement operator. Any attempt to modify the range of accessible pure steady states would then require engineering a new measurement scheme on the system. The ability to do this, however, depends on the specific details of the model and is subject to the limitations imposed by the physical system analysed. Here we have shown that, for a specific measurement scheme, quantum-jump-based feedback can be used to alter the dynamics of the system and the stability of steady states. In the future, it would be interesting to consider new measurement procedures that, allied to feedback design, would open up the possibilities to engineer a broader range of interesting multipartite steady states.

\bibliography{$HOME/ARTICLES/allbib}

\begin{thebibliography}{42}
\expandafter\ifx\csname natexlab\endcsname\relax\def\natexlab#1{#1}\fi
\expandafter\ifx\csname bibnamefont\endcsname\relax
  \def\bibnamefont#1{#1}\fi
\expandafter\ifx\csname bibfnamefont\endcsname\relax
  \def\bibfnamefont#1{#1}\fi
\expandafter\ifx\csname citenamefont\endcsname\relax
  \def\citenamefont#1{#1}\fi
\expandafter\ifx\csname url\endcsname\relax
  \def\url#1{\texttt{#1}}\fi
\expandafter\ifx\csname urlprefix\endcsname\relax\def\urlprefix{URL }\fi
\providecommand{\bibinfo}[2]{#2}
\providecommand{\eprint}[2][]{\url{#2}}

\bibitem[{\citenamefont{Poyatos et~al.}(1996)\citenamefont{Poyatos, Cirac, and
  Zoller}}]{Poyatos:1996}
\bibinfo{author}{\bibfnamefont{J.~F.} \bibnamefont{Poyatos}},
  \bibinfo{author}{\bibfnamefont{J.~I.} \bibnamefont{Cirac}}, \bibnamefont{and}
  \bibinfo{author}{\bibfnamefont{P.}~\bibnamefont{Zoller}},
  \bibinfo{journal}{Phys. Rev. Lett.} \textbf{\bibinfo{volume}{77}},
  \bibinfo{pages}{4728} (\bibinfo{year}{1996}).

\bibitem[{\citenamefont{Carvalho et~al.}(2001)\citenamefont{Carvalho, Milman,
  de~Matos~Filho, and Davidovich}}]{Carvalho:2001}
\bibinfo{author}{\bibfnamefont{A.~R.~R.} \bibnamefont{Carvalho}},
  \bibinfo{author}{\bibfnamefont{P.}~\bibnamefont{Milman}},
  \bibinfo{author}{\bibfnamefont{R.~L.} \bibnamefont{de~Matos~Filho}},
  \bibnamefont{and}
  \bibinfo{author}{\bibfnamefont{L.}~\bibnamefont{Davidovich}},
  \bibinfo{journal}{Phys. Rev. Lett.} \textbf{\bibinfo{volume}{86}},
  \bibinfo{pages}{4988} (\bibinfo{year}{2001}).

\bibitem[{\citenamefont{Viola et~al.}(1999)\citenamefont{Viola, Knill, and
  Lloyd}}]{Viola:1999}
\bibinfo{author}{\bibfnamefont{L.}~\bibnamefont{Viola}},
  \bibinfo{author}{\bibfnamefont{E.}~\bibnamefont{Knill}}, \bibnamefont{and}
  \bibinfo{author}{\bibfnamefont{S.}~\bibnamefont{Lloyd}},
  \bibinfo{journal}{Phys. Rev. Lett.} \textbf{\bibinfo{volume}{82}},
  \bibinfo{pages}{2417} (\bibinfo{year}{1999}).

\bibitem[{\citenamefont{Viola}(2004)}]{Viola:2004}
\bibinfo{author}{\bibfnamefont{L.}~\bibnamefont{Viola}},
  \bibinfo{journal}{Journal of Modern Optics} \textbf{\bibinfo{volume}{51}},
  \bibinfo{pages}{2357} (\bibinfo{year}{2004}).

\bibitem[{\citenamefont{Zanardi and Rasetti}(1997)}]{Zanardi:1997}
\bibinfo{author}{\bibfnamefont{P.}~\bibnamefont{Zanardi}} \bibnamefont{and}
  \bibinfo{author}{\bibfnamefont{M.}~\bibnamefont{Rasetti}},
  \bibinfo{journal}{Phys. Rev. Lett.} \textbf{\bibinfo{volume}{79}},
  \bibinfo{pages}{3306} (\bibinfo{year}{1997}).

\bibitem[{\citenamefont{Lidar et~al.}(1998)\citenamefont{Lidar, Chuang, and
  Whaley}}]{Lidar:1998}
\bibinfo{author}{\bibfnamefont{D.~A.} \bibnamefont{Lidar}},
  \bibinfo{author}{\bibfnamefont{I.~L.} \bibnamefont{Chuang}},
  \bibnamefont{and} \bibinfo{author}{\bibfnamefont{K.~B.}
  \bibnamefont{Whaley}}, \bibinfo{journal}{Phys. Rev. Lett.}
  \textbf{\bibinfo{volume}{81}}, \bibinfo{pages}{2594} (\bibinfo{year}{1998}).

\bibitem[{\citenamefont{Warren et~al.}(1993)\citenamefont{Warren, Rabitz, and
  Dahleh}}]{Warren:1993}
\bibinfo{author}{\bibfnamefont{W.~S.} \bibnamefont{Warren}},
  \bibinfo{author}{\bibfnamefont{H.}~\bibnamefont{Rabitz}}, \bibnamefont{and}
  \bibinfo{author}{\bibfnamefont{M.}~\bibnamefont{Dahleh}},
  \bibinfo{journal}{Science} \textbf{\bibinfo{volume}{259}},
  \bibinfo{pages}{1581} (\bibinfo{year}{1993}),
  \urlprefix\url{http://www.sciencemag.org/cgi/content/abstract/259/5101/1581}.

\bibitem[{\citenamefont{Wiseman}(1994)}]{Wiseman:1994}
\bibinfo{author}{\bibfnamefont{H.~M.} \bibnamefont{Wiseman}},
  \bibinfo{journal}{Phys. Rev. A} \textbf{\bibinfo{volume}{49}},
  \bibinfo{pages}{2133} (\bibinfo{year}{1994}).

\bibitem[{\citenamefont{Vitali et~al.}(1998)\citenamefont{Vitali, Tombesi, and
  Milburn}}]{Vitali:1998}
\bibinfo{author}{\bibfnamefont{D.}~\bibnamefont{Vitali}},
  \bibinfo{author}{\bibfnamefont{P.}~\bibnamefont{Tombesi}}, \bibnamefont{and}
  \bibinfo{author}{\bibfnamefont{G.~J.} \bibnamefont{Milburn}},
  \bibinfo{journal}{Phys. Rev. A} \textbf{\bibinfo{volume}{57}},
  \bibinfo{pages}{4930} (\bibinfo{year}{1998}).

\bibitem[{\citenamefont{{Doherty} and {Jacobs}}(1999)}]{Doherty:1999}
\bibinfo{author}{\bibfnamefont{A.~C.} \bibnamefont{{Doherty}}}
  \bibnamefont{and} \bibinfo{author}{\bibfnamefont{K.}~\bibnamefont{{Jacobs}}},
  \bibinfo{journal}{Phys. Rev. A} \textbf{\bibinfo{volume}{60}},
  \bibinfo{pages}{2700} (\bibinfo{year}{1999}).

\bibitem[{\citenamefont{Verstraete et~al.}(2009)\citenamefont{Verstraete, Wolf,
  and Ignacio~Cirac}}]{Verstraete:2009}
\bibinfo{author}{\bibfnamefont{F.}~\bibnamefont{Verstraete}},
  \bibinfo{author}{\bibfnamefont{M.~M.} \bibnamefont{Wolf}}, \bibnamefont{and}
  \bibinfo{author}{\bibfnamefont{J.}~\bibnamefont{Ignacio~Cirac}},
  \bibinfo{journal}{Nat Phys} \textbf{\bibinfo{volume}{5}},
  \bibinfo{pages}{633} (\bibinfo{year}{2009}),
  \urlprefix\url{http://dx.doi.org/10.1038/nphys1342}.

\bibitem[{\citenamefont{Mancini and Wang}(2005)}]{Mancini:2005}
\bibinfo{author}{\bibfnamefont{S.}~\bibnamefont{Mancini}} \bibnamefont{and}
  \bibinfo{author}{\bibfnamefont{J.}~\bibnamefont{Wang}},
  \bibinfo{journal}{Eur. Phys. J. D} \textbf{\bibinfo{volume}{32}},
  \bibinfo{pages}{257} (\bibinfo{year}{2005}).

\bibitem[{\citenamefont{Mancini and Wiseman}(2007)}]{Mancini:2007}
\bibinfo{author}{\bibfnamefont{S.}~\bibnamefont{Mancini}} \bibnamefont{and}
  \bibinfo{author}{\bibfnamefont{H.~M.} \bibnamefont{Wiseman}},
  \bibinfo{journal}{Physical Review A} \textbf{\bibinfo{volume}{75}},
  \bibinfo{eid}{012330} (pages~\bibinfo{numpages}{10}) (\bibinfo{year}{2007}),
  \urlprefix\url{http://link.aps.org/abstract/PRA/v75/e012330}.

\bibitem[{\citenamefont{Stockton et~al.}(2004)\citenamefont{Stockton, van
  Handel, and Mabuchi}}]{Stockton:2004}
\bibinfo{author}{\bibfnamefont{J.~K.} \bibnamefont{Stockton}},
  \bibinfo{author}{\bibfnamefont{R.}~\bibnamefont{van Handel}},
  \bibnamefont{and} \bibinfo{author}{\bibfnamefont{H.}~\bibnamefont{Mabuchi}},
  \bibinfo{journal}{Phys. Rev. A} \textbf{\bibinfo{volume}{70}},
  \bibinfo{pages}{022106} (\bibinfo{year}{2004}).

\bibitem[{\citenamefont{Wang et~al.}(2005)\citenamefont{Wang, Wiseman, and
  Milburn}}]{Wang:2005}
\bibinfo{author}{\bibfnamefont{J.}~\bibnamefont{Wang}},
  \bibinfo{author}{\bibfnamefont{H.~M.} \bibnamefont{Wiseman}},
  \bibnamefont{and} \bibinfo{author}{\bibfnamefont{G.~J.}
  \bibnamefont{Milburn}}, \bibinfo{journal}{Phys. Rev. A}
  \textbf{\bibinfo{volume}{71}}, \bibinfo{pages}{042309}
  (\bibinfo{year}{2005}).

\bibitem[{\citenamefont{Carvalho and Hope}(2007)}]{Carvalho:2007}
\bibinfo{author}{\bibfnamefont{A.~R.~R.} \bibnamefont{Carvalho}}
  \bibnamefont{and} \bibinfo{author}{\bibfnamefont{J.~J.} \bibnamefont{Hope}},
  \bibinfo{journal}{Phys. Rev. A} \textbf{\bibinfo{volume}{76}},
  \bibinfo{pages}{010301(R)} (\bibinfo{year}{2007}).

\bibitem[{\citenamefont{Carvalho et~al.}(2008)\citenamefont{Carvalho, Reid, and
  Hope}}]{Carvalho:2008}
\bibinfo{author}{\bibfnamefont{A.~R.~R.} \bibnamefont{Carvalho}},
  \bibinfo{author}{\bibfnamefont{A.~J.~S.} \bibnamefont{Reid}},
  \bibnamefont{and} \bibinfo{author}{\bibfnamefont{J.~J.} \bibnamefont{Hope}},
  \bibinfo{journal}{Phys. Rev. A} \textbf{\bibinfo{volume}{78}},
  \bibinfo{pages}{012334} (\bibinfo{year}{2008}).

\bibitem[{\citenamefont{Stevenson et~al.}(2011)\citenamefont{Stevenson,
  Carvalho, and Hope}}]{Stevenson:2011}
\bibinfo{author}{\bibfnamefont{R.~N.} \bibnamefont{Stevenson}},
  \bibinfo{author}{\bibfnamefont{A.~R.~R.} \bibnamefont{Carvalho}},
  \bibnamefont{and} \bibinfo{author}{\bibfnamefont{J.~J.} \bibnamefont{Hope}},
  \bibinfo{journal}{Eur. Phys. J. D} \textbf{\bibinfo{volume}{61}},
  \bibinfo{pages}{523} (\bibinfo{year}{2011}),
  \urlprefix\url{http://dx.doi.org/10.1140/epjd/e2010-10442-2}.

\bibitem[{\citenamefont{Carvalho and Santos}(2011)}]{Carvalho:2011}
\bibinfo{author}{\bibfnamefont{A.~R.~R.} \bibnamefont{Carvalho}}
  \bibnamefont{and} \bibinfo{author}{\bibfnamefont{M.~F.}
  \bibnamefont{Santos}}, \bibinfo{journal}{New Journal of Physics}
  \textbf{\bibinfo{volume}{13}}, \bibinfo{pages}{013010}
  (\bibinfo{year}{2011}),
  \urlprefix\url{http://stacks.iop.org/1367-2630/13/i=1/a=013010}.

\bibitem[{\citenamefont{Roos et~al.}(2004)\citenamefont{Roos, Lancaster, Riebe,
  H\"affner, H\"ansel, Gulde, Becher, Eschner, Schmidt-Kaler, and
  Blatt}}]{Roos:2004}
\bibinfo{author}{\bibfnamefont{C.~F.} \bibnamefont{Roos}},
  \bibinfo{author}{\bibfnamefont{G.~P.~T.} \bibnamefont{Lancaster}},
  \bibinfo{author}{\bibfnamefont{M.}~\bibnamefont{Riebe}},
  \bibinfo{author}{\bibfnamefont{H.}~\bibnamefont{H\"affner}},
  \bibinfo{author}{\bibfnamefont{W.}~\bibnamefont{H\"ansel}},
  \bibinfo{author}{\bibfnamefont{S.}~\bibnamefont{Gulde}},
  \bibinfo{author}{\bibfnamefont{C.}~\bibnamefont{Becher}},
  \bibinfo{author}{\bibfnamefont{J.}~\bibnamefont{Eschner}},
  \bibinfo{author}{\bibfnamefont{F.}~\bibnamefont{Schmidt-Kaler}},
  \bibnamefont{and} \bibinfo{author}{\bibfnamefont{R.}~\bibnamefont{Blatt}},
  \bibinfo{journal}{Phys. Rev. Lett.} \textbf{\bibinfo{volume}{92}},
  \bibinfo{pages}{220402} (\bibinfo{year}{2004}).

\bibitem[{\citenamefont{H{\"a}ffner et~al.}(2005)\citenamefont{H{\"a}ffner,
  Schmidt-Kaler, H{\"a}nsel, Roos, K{\"o}rber, Chwalla, Riebe, Benhelm, Rapol,
  Becher et~al.}}]{Haffner:2005a}
\bibinfo{author}{\bibfnamefont{H.}~\bibnamefont{H{\"a}ffner}},
  \bibinfo{author}{\bibfnamefont{F.}~\bibnamefont{Schmidt-Kaler}},
  \bibinfo{author}{\bibfnamefont{W.}~\bibnamefont{H{\"a}nsel}},
  \bibinfo{author}{\bibfnamefont{C.~F.} \bibnamefont{Roos}},
  \bibinfo{author}{\bibfnamefont{T.}~\bibnamefont{K{\"o}rber}},
  \bibinfo{author}{\bibfnamefont{M.}~\bibnamefont{Chwalla}},
  \bibinfo{author}{\bibfnamefont{M.}~\bibnamefont{Riebe}},
  \bibinfo{author}{\bibfnamefont{J.}~\bibnamefont{Benhelm}},
  \bibinfo{author}{\bibfnamefont{U.~D.} \bibnamefont{Rapol}},
  \bibinfo{author}{\bibfnamefont{C.}~\bibnamefont{Becher}},
  \bibnamefont{et~al.}, \bibinfo{journal}{Applied Physics B: Lasers and Optics}
  \textbf{\bibinfo{volume}{81}}, \bibinfo{pages}{151} (\bibinfo{year}{2005}).

\bibitem[{\citenamefont{Carvalho et~al.}(2004)\citenamefont{Carvalho, Mintert,
  and Buchleitner}}]{Carvalho:2004a}
\bibinfo{author}{\bibfnamefont{A.~R.~R.} \bibnamefont{Carvalho}},
  \bibinfo{author}{\bibfnamefont{F.}~\bibnamefont{Mintert}}, \bibnamefont{and}
  \bibinfo{author}{\bibfnamefont{A.}~\bibnamefont{Buchleitner}},
  \bibinfo{journal}{Phys. Rev. Lett.} \textbf{\bibinfo{volume}{93}},
  \bibinfo{pages}{230501} (\bibinfo{year}{2004}).

\bibitem[{\citenamefont{Simon and Kempe}(2002)}]{Simon:2002}
\bibinfo{author}{\bibfnamefont{C.}~\bibnamefont{Simon}} \bibnamefont{and}
  \bibinfo{author}{\bibfnamefont{J.}~\bibnamefont{Kempe}},
  \bibinfo{journal}{Phys. Rev. A} \textbf{\bibinfo{volume}{65}},
  \bibinfo{pages}{052327} (\bibinfo{year}{2002}).

\bibitem[{\citenamefont{D\"ur and Briegel}(2004)}]{Dur:2004}
\bibinfo{author}{\bibfnamefont{W.}~\bibnamefont{D\"ur}} \bibnamefont{and}
  \bibinfo{author}{\bibfnamefont{H.-J.} \bibnamefont{Briegel}},
  \bibinfo{journal}{Phys. Rev. Lett.} \textbf{\bibinfo{volume}{92}},
  \bibinfo{pages}{180403} (\bibinfo{year}{2004}).

\bibitem[{\citenamefont{Lindblad}(1976)}]{Lindblad:1976}
\bibinfo{author}{\bibfnamefont{G.}~\bibnamefont{Lindblad}},
  \bibinfo{journal}{Math. Phys.} \textbf{\bibinfo{volume}{48}},
  \bibinfo{pages}{119} (\bibinfo{year}{1976}).

\bibitem[{\citenamefont{Gorini et~al.}(1976)\citenamefont{Gorini, Kossakowski,
  and Sudarshan}}]{Gorini:1976}
\bibinfo{author}{\bibfnamefont{V.}~\bibnamefont{Gorini}},
  \bibinfo{author}{\bibfnamefont{A.}~\bibnamefont{Kossakowski}},
  \bibnamefont{and} \bibinfo{author}{\bibfnamefont{E.~C.~G.}
  \bibnamefont{Sudarshan}}, \bibinfo{journal}{J. Math. Phys.}
  \textbf{\bibinfo{volume}{17}}, \bibinfo{pages}{821} (\bibinfo{year}{1976}).

\bibitem[{\citenamefont{Carmichael}(1993)}]{Carmichael:1993}
\bibinfo{author}{\bibfnamefont{H.}~\bibnamefont{Carmichael}},
  \emph{\bibinfo{title}{An open systems approach to quantum optics}}, Lecture
  Notes in Physics m 18 (\bibinfo{publisher}{Springer-Verlag, Berlin},
  \bibinfo{year}{1993}).

\bibitem[{\citenamefont{M{\o}lmer et~al.}(1993)\citenamefont{M{\o}lmer, Castin,
  and Dalibard}}]{Molmer:1993}
\bibinfo{author}{\bibfnamefont{K.}~\bibnamefont{M{\o}lmer}},
  \bibinfo{author}{\bibfnamefont{Y.}~\bibnamefont{Castin}}, \bibnamefont{and}
  \bibinfo{author}{\bibfnamefont{J.}~\bibnamefont{Dalibard}},
  \bibinfo{journal}{J. Opt. Soc. Am. B} \textbf{\bibinfo{volume}{10}},
  \bibinfo{pages}{524} (\bibinfo{year}{1993}).

\bibitem[{\citenamefont{Spohn}(1976)}]{Spohn:1976}
\bibinfo{author}{\bibfnamefont{H.}~\bibnamefont{Spohn}},
  \bibinfo{journal}{Reports on Mathematical Physics}
  \textbf{\bibinfo{volume}{10}}, \bibinfo{pages}{189 } (\bibinfo{year}{1976}),
  ISSN \bibinfo{issn}{0034-4877},
  \urlprefix\url{http://www.sciencedirect.com/science/article/B6VN0-45FSNFG-6/%
2/99922863bab7d6822175316d00249135}.

\bibitem[{\citenamefont{Spohn}(1977)}]{Spohn:1977}
\bibinfo{author}{\bibfnamefont{H.}~\bibnamefont{Spohn}},
  \bibinfo{journal}{Letters in Mathematical Physics}
  \textbf{\bibinfo{volume}{2}}, \bibinfo{pages}{33} (\bibinfo{year}{1977}),
  ISSN \bibinfo{issn}{0377-9017}, \bibinfo{note}{10.1007/BF00420668},
  \urlprefix\url{http://dx.doi.org/10.1007/BF00420668}.

\bibitem[{\citenamefont{Frigerio}(1978)}]{Frigerio:1978}
\bibinfo{author}{\bibfnamefont{A.}~\bibnamefont{Frigerio}},
  \bibinfo{journal}{Comm. Math. Phys.} \textbf{\bibinfo{volume}{63}},
  \bibinfo{pages}{269} (\bibinfo{year}{1978}).

\bibitem[{\citenamefont{Schirmer and Wang}(2010)}]{Schirmer:2010}
\bibinfo{author}{\bibfnamefont{S.~G.} \bibnamefont{Schirmer}} \bibnamefont{and}
  \bibinfo{author}{\bibfnamefont{X.}~\bibnamefont{Wang}},
  \bibinfo{journal}{Phys. Rev. A} \textbf{\bibinfo{volume}{81}},
  \bibinfo{pages}{062306} (\bibinfo{year}{2010}).

\bibitem[{\citenamefont{T\'oth et~al.}(2007)\citenamefont{T\'oth, Knapp,
  G\"uhne, and Briegel}}]{Toth:2007}
\bibinfo{author}{\bibfnamefont{G.}~\bibnamefont{T\'oth}},
  \bibinfo{author}{\bibfnamefont{C.}~\bibnamefont{Knapp}},
  \bibinfo{author}{\bibfnamefont{O.}~\bibnamefont{G\"uhne}}, \bibnamefont{and}
  \bibinfo{author}{\bibfnamefont{H.~J.} \bibnamefont{Briegel}},
  \bibinfo{journal}{Phys. Rev. Lett.} \textbf{\bibinfo{volume}{99}},
  \bibinfo{pages}{250405} (\bibinfo{year}{2007}).

\bibitem[{\citenamefont{Livine and Terno}(2005)}]{Livine:2005}
\bibinfo{author}{\bibfnamefont{E.~R.} \bibnamefont{Livine}} \bibnamefont{and}
  \bibinfo{author}{\bibfnamefont{D.~R.} \bibnamefont{Terno}},
  \bibinfo{journal}{Phys. Rev. A} \textbf{\bibinfo{volume}{72}},
  \bibinfo{pages}{022307} (\bibinfo{year}{2005}).

\bibitem[{\citenamefont{Yao}(2011)}]{Yao:2011}
\bibinfo{author}{\bibfnamefont{W.}~\bibnamefont{Yao}},
  \bibinfo{journal}{arXiv:1101.3888v1}  (\bibinfo{year}{2011}).

\bibitem[{\citenamefont{T\'oth and Mitchell}(2010)}]{Toth:2010}
\bibinfo{author}{\bibfnamefont{G.}~\bibnamefont{T\'oth}} \bibnamefont{and}
  \bibinfo{author}{\bibfnamefont{M.~W.} \bibnamefont{Mitchell}},
  \bibinfo{journal}{New Journal of Physics} \textbf{\bibinfo{volume}{12}},
  \bibinfo{pages}{053007} (\bibinfo{year}{2010}),
  \urlprefix\url{http://stacks.iop.org/1367-2630/12/i=5/a=053007}.

\bibitem[{\citenamefont{Bartlett et~al.}(2003)\citenamefont{Bartlett, Rudolph,
  and Spekkens}}]{Bartlett:2003}
\bibinfo{author}{\bibfnamefont{S.~D.} \bibnamefont{Bartlett}},
  \bibinfo{author}{\bibfnamefont{T.}~\bibnamefont{Rudolph}}, \bibnamefont{and}
  \bibinfo{author}{\bibfnamefont{R.~W.} \bibnamefont{Spekkens}},
  \bibinfo{journal}{Phys. Rev. Lett.} \textbf{\bibinfo{volume}{91}},
  \bibinfo{pages}{027901} (\bibinfo{year}{2003}).

\bibitem[{\citenamefont{Cabello}(2003)}]{Cabello:2003}
\bibinfo{author}{\bibfnamefont{A.}~\bibnamefont{Cabello}}, \bibinfo{journal}{J.
  Mod. Opt} \textbf{\bibinfo{volume}{50}}, \bibinfo{pages}{1049}
  (\bibinfo{year}{2003}).

\bibitem[{\citenamefont{Mintert and Buchleitner}(2005)}]{Mintert:2005}
\bibinfo{author}{\bibfnamefont{F.}~\bibnamefont{Mintert}} \bibnamefont{and}
  \bibinfo{author}{\bibfnamefont{A.}~\bibnamefont{Buchleitner}},
  \bibinfo{journal}{Phys. Rev. A} \textbf{\bibinfo{volume}{72}},
  \bibinfo{pages}{012336} (\bibinfo{year}{2005}).

\bibitem[{\citenamefont{Greenberger et~al.}(1989)\citenamefont{Greenberger,
  Horne, and Zeilinger}}]{Greenberger:1989}
\bibinfo{author}{\bibfnamefont{D.~M.} \bibnamefont{Greenberger}},
  \bibinfo{author}{\bibfnamefont{M.~A.} \bibnamefont{Horne}}, \bibnamefont{and}
  \bibinfo{author}{\bibfnamefont{A.}~\bibnamefont{Zeilinger}}, in
  \emph{\bibinfo{booktitle}{Bell's Theorem, Quantum Theory and Conceptions of
  the Universe Bell's Theorem, Quantum Theory and Conceptions of the
  Universe}}, edited by
  \bibinfo{editor}{\bibfnamefont{M.}~\bibnamefont{Kafatos}}
  (\bibinfo{publisher}{Kluwer, Dodrecht}, \bibinfo{year}{1989}), pp.
  \bibinfo{pages}{73--76}.

\bibitem[{\citenamefont{D{\"u}r et~al.}(2000)\citenamefont{D{\"u}r, Vidal, and
  Cirac}}]{Dur:2000a}
\bibinfo{author}{\bibfnamefont{W.}~\bibnamefont{D{\"u}r}},
  \bibinfo{author}{\bibfnamefont{G.}~\bibnamefont{Vidal}}, \bibnamefont{and}
  \bibinfo{author}{\bibfnamefont{J.~I.} \bibnamefont{Cirac}},
  \bibinfo{journal}{Phys. Rev. A} \textbf{\bibinfo{volume}{62}},
  \bibinfo{pages}{062314} (\bibinfo{year}{2000}).

\bibitem[{\citenamefont{Briegel and Raussendorf}(2001)}]{Briegel:2001}
\bibinfo{author}{\bibfnamefont{H.~J.} \bibnamefont{Briegel}} \bibnamefont{and}
  \bibinfo{author}{\bibfnamefont{R.}~\bibnamefont{Raussendorf}},
  \bibinfo{journal}{Phys. Rev. Lett.} \textbf{\bibinfo{volume}{86}},
  \bibinfo{pages}{910} (\bibinfo{year}{2001}).

\end{thebibliography}

\end{document}